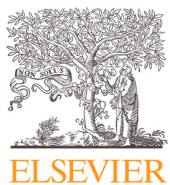
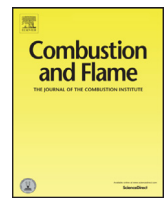

# A numerical study on combustion mode characterization for locally stratified dual-fuel mixtures

Shervin Karimkashi\*, Heikki Kahila, Ossi Kaario, Martti Larmi, Ville Vuorinen

*Department of Mechanical Engineering, School of Engineering, Aalto University, Otakaari 4, Espoo 02150, Finland*

**ABSTRACT**

Combustion modes in locally stratified dual-fuel (DF) mixtures are numerically investigated for methanol/n-dodecane blends under engine-relevant pressures. In the studied constant-volume numerical setup, methanol acts as a background low-reactivity fuel (LRF) while n-dodecane serves as high-reactivity fuel (HRF), controlling local ignition delay time. The spatial distribution of n-dodecane is modeled as a sinusoidal function parametrized by stratification amplitude ($Y'$) and wavelength (0.01 mm $< \lambda <$ 15 mm). In contrast, methanol is assumed to be fully premixed with air at equivalence ratio 0.8. First, one-dimensional setup is investigated by hundreds of chemical kinetics simulations in ($Y'$, $\lambda$) parameter space. Further, the concepts by Sankaran et al. (2005, *Proceedings of the Combustion Institute*) and Zeldovich (1980, *Combustion and Flame*) on ignition front propagation speed are applied to develop a theoretical analysis of the time-dependent diffusion–reaction problem. The theoretical analysis predicts two combustion modes, (1) spontaneous ignition and (2) deflagrative propagation, and leads to an analytical expression for the border curve called $\beta$-curve herein. One-dimensional chemical kinetics simulations confirm the presence of two combustion modes in ($Y'$, $\lambda$) parameter space while the $\beta$-curve explains consistently the position of phase border observed in the simulations. Finally, the role of convective mixing is incorporated to the theoretical expression for the $\beta$-curve. The effect of convection on combustion mode is assessed by carrying out two-dimensional fully-resolved simulations with different turbulence levels. Two-dimensional numerical simulation results give evidence on combustion mode switching, which is consistent with predictions of the modified $\beta$-curve for turbulent cases. The practical output of the paper is the $\beta$-curve which is proposed as a predictive tool to estimate combustion modes for various fuels or fuel combinations.

## 1. Introduction

Various environmental concerns on combustion related emissions and restricted resources of fossil fuels have inspired the utilization of sustainable fuels for both compression ignition (CI) and spark ignited (SI) engines. For example, with relevance to the present paper, methanol can be sustainably produced from various sources including solar and/or wind energy [1], and biogas [2,3].

Alcohol-containing fuels, such as methanol, have drawn attention of engine researchers in recent years mainly due to their renewability, wide source of availability, and high octane number, hence, knock mitigation [4,5]. Moreover, their cooling effect – due to their higher heat of vaporization compared to natural gas and gasoline – can reduce the charge peak combustion temperature which leads to NOx emission reduction [4,6]. Also, since such fuels contain oxygen and are free of aromatics, they mitigate soot formation [6].

However, alcohol-containing fuels such as methanol have relatively low energy density and low reactivity. For instance, energy density of methanol is about half compared to natural gas and gasoline [7]. Therefore, in CI context, they are commonly used in dual-fuel (DF) engines where a high reactivity pilot fuel, such as diesel, facilitates the ignition process. Dual-fuel CI strategies feature low NOx and soot formation, and high thermal efficiencies [8,9]. One example of DF engines is a so-called reactivity controlled compression ignition (RCCI) combustion strategy, where a blend of two fuels ignites in an interactive manner [10]. A common fuel injection strategy in RCCI context is to inject the low-reactivity fuel (LRF) to the intake manifold while injecting the high-reactivity fuel (HRF) directly to the cylinder as a liquid fuel spray with a relatively long premixing time, i.e. early injection. There are several studies on RCCI methanol-diesel combustion, e.g., [11–19]. For instance, Zou et al. [17] used different LRFs – including methanol, ethanol, n-butanol and gasoline – and diesel fuel as the HRF in numerical investigations of RCCI. They observed that with alcohol-containing fuels, combustion durations are shorter and pressure rise rates are higher than those with n-butanol and gasoline.

\* Corresponding author.
*E-mail address:* shervin.karimkashiarani@aalto.fi (S. Karimkashi).

In practice, RCCI engines operate under stratified conditions with local HRF gradients. Under such circumstances, the dominant heat release rate (HRR) mode has not been properly characterized. Many authors have considered RCCI combustion as a flameless mode with volumetric heat release, e.g., [11,16,20]. However, Zhou et al. [21] indicated a situation where spontaneous ignition and flame fronts were both present. Zeldovich has [22] presented the theoretical basis for classification of different regimes of ignition propagation and identified the following modes of combustion sorted from the highest (lowest) to the lowest (highest) combustion wave speed (temperature gradient): (i) instantaneous thermal explosion, (ii) autoignitive spontaneous ignition fronts propagating at speeds greater than normal detonation rate, (iii) subsonic spontaneous ignition front propagation, and (iv) normal deflagration. Amongst the different combustion modes, normal deflagration (diffusion important for HRR) and subsonic spontaneous ignition (diffusion less important for HRR) are favorable modes in engines. To characterize the dominant mode of combustion, Sankaran et al. [23] proposed a non-dimensional number based on the laminar flame speed and ignition speed to distinguish between normal deflagration and subsonic spontaneous propagation. Introduction of flame index [24], autoignition index [25], and chemical explosive mode analysis [26] are instances of other pathways taken by researchers to distinguish between different regimes of propagation.

In single-fuel (SF) constant-volume conditions, there is a body of literature available that scrutinize the importance of ignition and subsequent combustion modes, where effects of temperature [23,27–39], and/or composition [33–35,39,40], and/or velocity [28,31,35,36,38] fluctuations are taken into account. For instance, Pal et al. [28,38] proposed regime diagrams which specify different modes of combustion with velocity and temperature fluctuations. More recently, Luong et al. [39] studied the role of temperature and composition fluctuations based on a theoretical analysis for dimethyl ether/air mixtures and validated it against two-dimensional (2d) direct numerical simulations (DNS).

In DF combustion, for example in port-injected RCCI configuration, the primary LRF is nearly premixed while the HRF may be more stratified if injected as an HRF liquid spray. In such a dual-reactivity scenario, Luong et al. [21] studied numerically the formation of autoignition kernels – driven by chemical reactions and ignition delay time (IDT) gradients. They reported that autoignition kernels may develop to deflagrative flame fronts – driven additionally by diffusive and convective effects. Bhagatwala et al. [41] studied effects of temperature and fuel stratification on possible modes of combustion by performing one-dimensional (1d) and 2d DNS studies for n-heptane/iso-octane blends under RCCI conditions. However, the study was limited to specific initial conditions, including fixed turbulence and stratification levels. Other 2d DNS studies [42–44] highlighted influences of mixing time-scales, temperature, fuel stratification, and turbulence on ignition characteristics, and reported existence of both deflagrative and spontaneous ignition, specifically, at early phases of ignition. More recently, Liu et al. [45] carried out an experimental study to examine effects of different fuel stratification levels on the ratio of flame front propagation and autoignition in RCCI, where increased port injection mass fractions resulted in observable flame front propagation.

Based on the presented literature survey, it is known that DF mixtures may either ignite volumetrically or develop a deflagrative flame front after the initial ignition. However, a systematic method to examine prevalence of deflagration versus spontaneous combustion under different HRF stratification levels in dual-reactivity mixtures is still lacking. According to the identified knowledge gaps, this study seeks to: (i) present a theoretical analysis to predict modes of DF combustion in engine-relevant conditions, (ii) verify the theoretical analysis by 1d reacting flow simulations to characterize combustion modes for methanol/n-dodecane DF combustion, and (iii) incorporate effects of turbulence to the presented analytical expression, and elaborate on the role of convective mixing by 2d fully-resolved simulations.

The paper is structured as follows. First, numerical solvers are introduced and a chemical mechanism is selected for the numerical simulations. Second, a discussion on characteristics of methanol/n-dodecane DF ignition using zero-dimensional (0d) studies is provided. Third, the theoretical analysis is presented. Fourth, the results of 1d chemical kinetics simulations are shown. Finally, effects of turbulence are incorporated to the theory and convection effects are explored using fully-resolved 2d numerical studies.

## 2. Numerical methods

### 2.1. Reactive flow solvers

In this work, two solvers are utilized for numerical computations; an open-source chemical kinetics library, Cantera [46], and a computational fluid dynamics (CFD) library, OpenFOAM [47]. Cantera is used only in Section 2.2 and Appendix A for validation of the utilized chemical mechanism, in Section 3.1 for 0d analysis of homogeneous mixtures, and to calculate IDT and laminar flame properties provided in Table 2 as model input parameters.

OpenFOAM is used to study ignition and subsequent deflagration in transient, stratified DF mixtures in the 1d and 2d chemical kinetics simulations. The compressible Navier-Stokes equations along with the transport equations for species mass fractions are solved using the direct integration of finite-rate chemistry within OpenFOAM 2.4.x framework. The customized solver is based on the recent work by Kahila et al. [48] where the open source library, pyJac [49], was coupled with OpenFOAM. PyJac provides reaction rate coefficients as well as the analytical Jacobian matrix formulation required by the ordinary differential equation (ODE) system solver. The chemistry solution is advanced within the operator splitting framework. The compressible solver utilizes the standard PISO (pressure implicit with splitting of operator) pressure-correction approach. The interested reader is referred to [48] for further details on the implementation and other improvements of the present approach over the standard OpenFOAM solvers, including enhanced ODE linear algebra and parallelization capabilities. It should be noted that this customized solver assumes infinite pressure wave speed and it has been shown to capture the compression effects after the first ignition [50].

In the present study, IDT is considered as the time instant when $T_{max} > 1200$ K, $T_{max}$ being the instantaneous maximum temperature. This definition has been previously used for DF ignition e.g., in Ref. [51]. Laminar flame thickness ($\delta_f$) is calculated based on the maximum gradient of temperature in the steady-state solution of 1d laminar premixed flame, $\delta_f = (T_b - T_u)/(\frac{\partial T}{\partial x})_{max}$, where $T_b$ and $T_u$ are burned (maximum) and unburned (minimum) temperatures and $x$ represents location along the simulation direction. Also, laminar flame speed ($S_l$) is equivalent to the reactants inflow speed in the steady-state solution of the 1d laminar premixed flame.

*2.2. Validation of chemical kinetics mechanism*

In this study, a diesel surrogate, n-dodecane (abbreviated as ndd), is chosen as the HRF and a renewable fuel, methanol, has been considered as the LRF in dual-reactivity simulations. We note that there is no specific chemical kinetics mechanism developed for n-dodecane/methanol/air DF combustion. Here, a reduced chemical mechanism developed for combustion of n-dodecane/air by Frassoldati et al. [52] is used. This mechanism is called Polimi reduced mechanism and it is validated using Cantera for SF methanol/air combustion in Appendix A for brevity. Previously, the mechanism was used and verified by the present authors for SF and DF studies of n-dodecane/methane ignition [48]. The validation in Appendix A shows that the Polimi reduced mechanism can capture ignition and flame properties of methanol/air combustion with a reliable precision and consequently, it is utilized in the present DF study.

## 3. Numerical setup and theoretical analysis

*3.1. 0d analysis of methanol/n-dodecane ignition*

As the main objective of the present study is to investigate modes of combustion in dual-reactivity conditions, a presentation of the IDT against mixture fraction ($0 < \xi < 1$) for methanol/air, n-dodecane/air, and n-dodecane/methanol-air mixtures gives a better insight to characteristics of ignition in DF combustion as compared with SF combustion. Mixture fraction describes the mixing extent of fuel ($\xi=1$) and oxidizer ($\xi=0$) in non-premixed combustion and here, it is defined based on the nitrogen mass fraction.

$$\xi = \frac{Y_{N_2} - Y_{N_2}^{ox}}{Y_{N_2}^{f} - Y_{N_2}^{ox}}, \quad (1)$$

where $Y_{N_2}$, $Y_{N_2}^{ox}$, and $Y_{N_2}^{f}$ represent the nitrogen mass fraction in the mixture, oxidizer stream and fuel stream (only n-dodecane in DF), respectively. The present definition is noted to be consistent with the conventional and Bilger's definition [53] for mixture fraction. As a remark, the Polimi reduced mechanism does not include $N_2$ related reactions. The same definition was previously utilized by Demosthenous et al. [51] for DF studies.

Inert mixing between the fuel and oxidizer streams is assumed in the 0d reactor model in Cantera where the Polimi reduced mechanism is utilized. At each mixture fraction, the mixture has a homogeneous temperature and composition which is calculated based on the inert mixing of fuel and oxidizer streams according to their enthalpies, $h = \sum_{\alpha=1}^{N} Y_\alpha h_\alpha$, where $h$, $Y$, $N$ and $\alpha$ are the absolute enthalpy, mass fraction, number of species and species index, respectively. The procedure was proposed in Refs. [54,55], and utilized also in Ref. [51]. For SF conditions, the fuel stream is either n-dodecane or methanol and the oxidizer stream is air. For the DF test, n-dodecane is considered as the fuel stream ($\xi = 1$), while the premixed blend of methanol/air represents the oxidizer stream ($\xi = 0$). The oxidizer consists of 21% $O_2$, 69% $N_2$, and 10% methanol where $\phi_{methanol} = 0.8$ ($\phi$ denoting the equivalence ratio). The gas composition has close correspondence with the simulations performed in Refs. [11,17]. We note that in this DF mixture, the inert $N_2$ is partially replaced by methanol in order to maintain $O_2$ concentration at the constant level of 21% in the mixture as in SF cases. This strategy is consistent with that applied in e.g., [48].

Next, two different scenarios for initial temperature of the streams are considered: different versus same temperatures for the two streams. Figure 1 illustrates differences between the two scenarios. In the first scenario, the fuel stream temperature (363 K), the oxidizer stream temperature (900 K) and the ambient pressure (50 bar) correspond to the Engine Combustion Network (ECN)

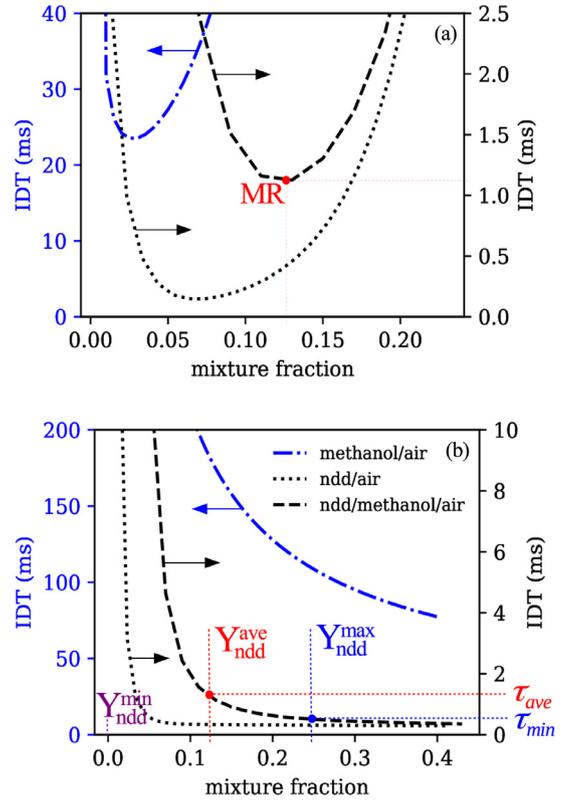

**Fig. 1.** Calculations of IDT versus mixture fraction using constant volume homogeneous reactor model at $p = 50$ bar, for three different mixtures: methanol/air, n-dodecane/air, and n-dodecane/methanol+air: (a) different initial temperatures for fuel (363 K) and oxidizer streams (900 K), (b) same initial temperatures for the two streams (770 K).

[56] Spray A experiments. In the second scenario, both streams are at the temperature of the most reactive mixture fraction of the DF case in the first scenario.

As marked in Fig. 1(a), the most reactive (MR) mixture fraction, $\xi_{MR}$, corresponds to the mixture with the shortest IDT, denoted as the most reactive IDT (IDT$_{MR}$). The corresponding temperature of this point (770 K) has been considered for initial temperature in both fuel and oxidizer streams at entire range of mixture fractions in the second scenario ($p = 50$ bar), presented in Fig. 1(b). For both scenarios, a significantly longer IDT$_{MR}$ for methanol compared with that of n-dodecane is observed, which is expected due to considerably higher reactivity of n-dodecane compared with methanol. In addition, $\xi_{MR}$ for methanol is considerably leaner than that of n-dodecane. For n-dodecane/methanol-air mixture, it is observed that methanol delays auto-ignition of n-dodecane. However, n-dodecane/methanol-air mixture exhibits a behavior closer to n-dodecane/air mixture rather than methanol/air.

In the present study, the second scenario (constant temperature) is investigated. As observed in Fig. 1(b), for both SF and DF combustion, IDT decreases at richer mixtures. When $\xi \to 0.3$, IDT of n-dodecane/methanol-air mixture approaches that of n-dodecane/air. It should be noted that in the 0d reactor model, an inert pre-mixing of n-dodecane/methanol-air is assumed, i.e. competition between mixing and reaction – as present in the following 1d and 2d case studies – is ignored. To set up the following 1d and 2d case studies, the average values of the mixture mass fractions are taken from Fig. 1(b) at $Y_{ndd}^{ave}$, such that the IDT is $\approx$ 1 ms. We note that here, the mixture fraction and n-dodecane mass fraction are equal. The maximum, the minimum and the average values of n-dodecane mass fraction ($Y_{ndd}$) and the corresponding IDT values

**Table 1**
Initial mass fractions for the highest stratification level in numerical simulations. All other cases with smaller amplitudes fluctuate around the average level ($Y_{ndd}^{ave}$, horizontal red line), while $\phi_{methanol}$ is kept at 0.8.

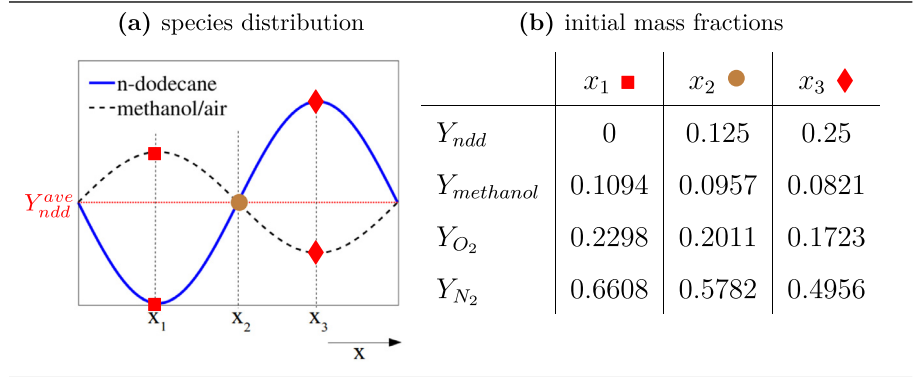

(a) species distribution    (b) initial mass fractions

|  | $x_1$ ■ | $x_2$ ● | $x_3$ ◆ |
|---|---|---|---|
| $Y_{ndd}$ | 0 | 0.125 | 0.25 |
| $Y_{methanol}$ | 0.1094 | 0.0957 | 0.0821 |
| $Y_{O_2}$ | 0.2298 | 0.2011 | 0.1723 |
| $Y_{N_2}$ | 0.6608 | 0.5782 | 0.4956 |

for the most stratified blend in the setup presented in the following section are marked in Fig. 1(b). These definitions are used later on in the paper.

### 3.2. Mixture initialization in 1d simulations

In the 1d numerical setup, we consider a premixed LRF charge at lean conditions ($\phi_{methanol} = 0.8$) where the HRF (n-dodecane) is distributed heterogeneously and temperature of the charge is considered to be constant at the time of ignition. The main idea in the initialization is to assume a constant equivalence ratio for LRF distribution, specify a simple HRF distribution in 1d, and then, let the mixture autoignite.

The dual-reactivity ignition of the stratified HRF-LRF mixture is specified in terms of two important parameters; $\lambda$ (wavelength) and $Y'(0)$ (amplitude). Here, we are interested in how mixtures ignite in the length-scales of 0.01 mm $< \lambda <$ 15 mm. The small-scale HRF distribution is described as a simple sinusoidal function,

$$Y_{ndd} = Y_{ndd}^{ave} + Y'(0)\sin(kx), \qquad (2)$$

where $Y_{ndd}$ represents n-dodecane mass fraction, $Y_{ndd}^{ave}$ is the average level of n-dodecane concentration in the mixture, $Y'(0)$ is the initial fluctuation of n-dodecane mass fraction, and $k$ is the wavenumber, $k = \frac{2\pi}{\lambda}$. When $Y'(0) \approx 0$, there is no stratification, and the system ignites spontaneously, similar to a homogeneous, constant volume reactor. On the other hand, considering the shorter IDT for richer mixture fraction values as noted in Fig. 1(b), when $Y'(0)$ is large, ignition starts from the richest locations evolving towards the leaner regions.

In this study, n-dodecane concentration fluctuates around an average mass fraction value for which the IDT of the mixture is around 1 ms, i.e. at $\xi = 0.125$ according to Fig. 1(b). The premixed LRF/air equivalence ratio is kept constant at $\phi_{methanol} = 0.8$ with a distribution that is reflection of the HRF distribution function over the x-axis with the same wavelength and its amplitude being $1 - Y'(0)$. Table 1 lists initial values of species mass fractions for the most heterogeneous case study. The maximum initial amplitude of n-dodecane fluctuations ($[Y'(0)]_{max}$) varies n-dodecane mass fraction between 0 and 0.25 and accordingly influences mass fractions of other species as reflected in Table 1.

Two different initial engine-relevant levels of pressure are considered. First, in case 1, initial temperature ($T_{init}$) is 770 K and pressure is set to 50 bar. This thermodynamic condition is taken according to the average values of the mixture mass fractions in Fig. 1(b) at $Y_{ndd}^{ave}$, such that the IDT is $\approx$ 1 ms. The second initial condition is found based on the non-reactive and isentropic pressure rise which leads to case 2 with $T_{init}$=850 K and pressure at 70 bar. We note that case 2 is only considered for exploring effects of pressure on the following analyses and its corresponding results are not provided with details for brevity. Table 2 presents information about 0d homogeneous reactor and 1d premixed flame calculations from Cantera for the aforementioned average initial conditions of species concentration, c.f. Table 1 (average values), and at corresponding initial thermodynamic conditions for case 1 and case 2.

**Table 2**
IDT values from 0d homogeneous reactor calculations and flame properties from 1d premixed flame calculations in Cantera for the average mixture composition considered in cases 1 and 2.

|  | $p$ (bar) | $T_{init}$ (K) | IDT (ms) | $S_l$ (m/s) | $\delta_f$ (mm) | $T_b$ (K) |
|---|---|---|---|---|---|---|
| Case 1 | 50 | 770 | 1.1 | 0.34 | 0.0177 | 1806.3 |
| Case 2 | 70 | 850 | 0.3 | 0.48 | 0.0112 | 1911.2 |

Based on the literature, steady flame fronts might not exist at such high pressure levels as considered herein [57]. However, for the selected thermochemical conditions in this paper, Cantera simulations converge enabling calculation of laminar flame speed and thickness. Moreover, Krisman et al. [57] compared DNS results with Cantera at high pressure and temperature conditions ($p = 40$ bar and 600 K $< T <$ 1400 K) and showed that Cantera results are reliable under those conditions. In the present modeling study we rely on the simulated values of $S_l$ and $\delta_f$.

IDT values of homogeneous reactor calculations in case 1 and case 2 – as reported in Table 2 – are denoted as $\tau_1$ and $\tau_2$, respectively. In this paper, case 1 is the major focus for theoretical and numerical analyses wherein $\tau_1$ is utilized to normalize time. For brevity, case 2 and its corresponding results are only discussed briefly in the text. It is noteworthy that for the stratified DF mixtures, $\tau_1$ is equivalent to IDT at $Y_{ndd}^{ave}$ ($\tau_{ave}$) in Fig. 1(b) and it is of the same order of magnitude as the IDT at $Y_{ndd}^{max}$ ($\tau_{min}$). In the following 1d numerical simulations, simulation times of 1.5 IDT are considered in order to ensure that the entire charge is combusted prior to the end of simulation.

With relevance to the second objective of this study, 300 1d simulations with different initial levels of n-dodecane stratification are carried out for both pressure levels (150 for each case). In the parameter sweeps, n-dodecane stratification level is introduced by varying the values of $Y'(0)$ and $\lambda$. The 1d constant-volume cases, with periodic boundary conditions, are initialized using sinusoidal profiles of species concentration (c.f. Eq. (2) and Table 1) with uniform temperature and pressure in OpenFOAM. Domain length is varied according to the wavelength such that one sinusoidal wave is present in the domain for each simulation, and grids are refined such that flame thickness is resolved by 20–40 gridpoints. We note that as periodic boundary conditions are considered, the systematic decrease of the domain size with the stratification wavelength is conceptually equivalent to having the domain length fixed and decreasing the wavelength. Simulation of only a single sinusoidal profile reduces the computational cost.

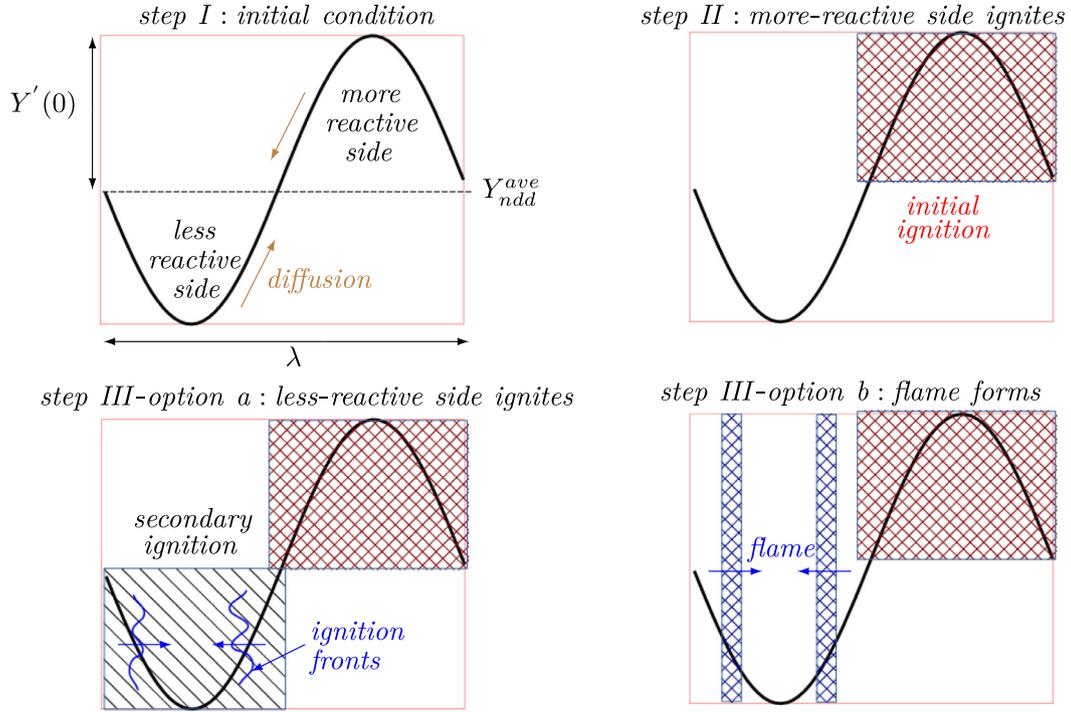

**Fig. 2.** A schematic representation for the two modes of combustion after initial ignition; spontaneous (step III-option a) versus deflagrative (option b).

### 3.3. Qualitative description of the 1d model dynamics

Considering the initialization process, the anticipated ignition phasing is schematically presented in Fig. 2. Due to the sinusoidal initialization of n-dodecane distribution in the system (step I), the more-reactive side of the domain with higher n-dodecane concentration ignites first (step II). During such an initial ignition, reactants on the less-reactive side are compressed and heated in the constant volume system. Consequently, this compression effect might disturb the sinusoidal curve and more importantly, it increases reactivity of remaining reactants. Two alternative options are expected in step III. In step III-a, due to either strong heat and mass diffusion or a small n-dodecane stratification amplitude, the less-reactive side of the domain ignites with a slight delay compared to the initial ignition, where ignition fronts form. On the other hand, in step III-b, flame fronts form due to the high IDT gradients in the system with relatively high stratification levels. The two fronts emanate from the initial ignition in the most reactive side in step II and they start to evolve symmetrically towards both the boundaries. Considering periodic boundary conditions, the two fronts appear as developed flame fronts propagating towards each other in the less reactive side. Next, these schematic steps are theoretically analyzed.

### 3.4. Theoretical analysis of the 1d model

For a homogeneous mixture with temperature fluctuations, Sankaran et al. [23] defined the parameter $\beta$, which specifies the relative dominance of deflagrative versus spontaneous combustion modes,

$$\beta = C_\beta \frac{S_l}{S_{ign}}, \quad (3)$$

where $C_\beta$ is a constant of order unity. Above, $S_l$ can be easily calculated by simulations (0.34 m/s for case 1), and $S_{ign}$ corresponds to the autoignition front propagation speed at time instants $t > $ IDT. Values of $\beta$ less (larger) than unity represent spontaneous ignition (deflagration) and $\beta = 1$ serves as the borderline between the two combustion modes. This idea is further developed here for a stratified mixture with homogeneous temperature in a simplified time-dependent diffusion–reaction problem.

According to Zeldovich [22], denoting IDT by $\tau$, $S_{ign}$ can be characterized by

$$S_{ign} = \frac{1}{|\nabla \tau|}. \quad (4)$$

For a 1d non-uniform composition and uniform temperature conditions, $\tau$ is a function of the composition fluctuations ($Y'$), and Eq. (4) can be rearranged as

$$\frac{1}{S_{ign}} = |\nabla \tau| = \frac{\partial \tau}{\partial x} = \underbrace{\frac{\partial \tau}{\partial Y'}}_{\text{I}} \underbrace{\frac{\partial Y'}{\partial x}}_{\text{II}}. \quad (5)$$

In Eq. (5), term I indicates variations of $\tau$ with $Y'$ which is specified by IDT vs $\xi$ plot in Fig. 1(b). We note that $\frac{\partial}{\partial Y'} = \frac{\partial}{\partial \xi}$. To simplify the theoretical analysis, we assume that $\tau$ decreases linearly when $\xi$ increases, and for case 1, the slope $\alpha$ ($\approx 0.2$ s) is defined by linearization according to Fig. 1(b),

$$\frac{\partial \tau}{\partial Y'} = \frac{\tau_{max} - \tau_{ave}}{Y_{ndd}^{ave} - Y_{ndd}^{min}} = \alpha. \quad (6)$$

It should be mentioned that when $Y_{ndd}^{min}$ tends to zero, $\tau_{max}$ approaches infinity. Accordingly, to find the linear correlation in Eq. (6), we take $Y_{ndd}^{min}$ as a small value, here 0.01, and based on that $\tau_{max} = 200$ ms according to Fig. 1(b).

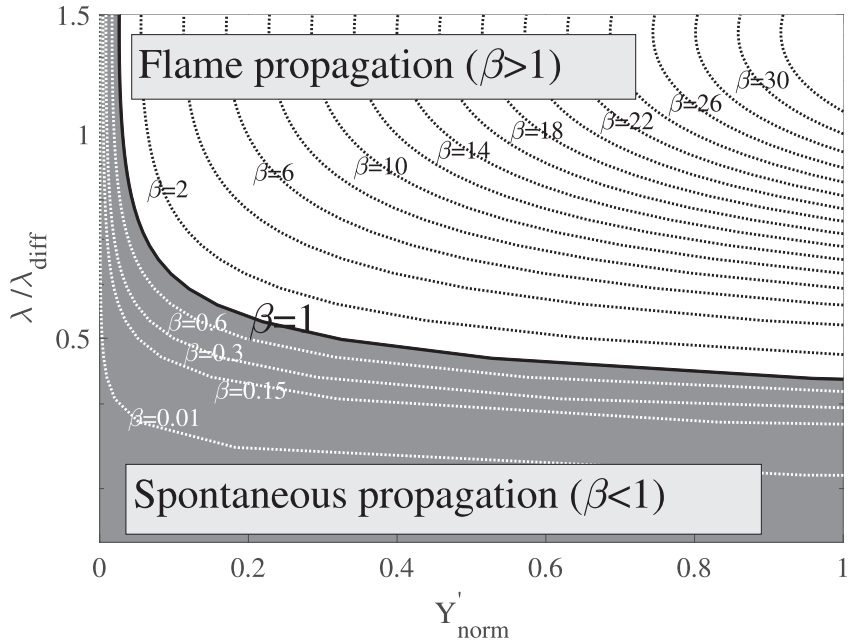

**Fig. 3.** Regime diagram based on the theoretical analysis. Eq. (13) is plotted for various values of $\beta$ ($S_l = 0.34$, $C_\beta = 1.0$). The solid line ($\beta$-curve) separates spontaneous (gray-shaded) and deflagrative (white region) combustion modes.

In Eq. (5), term II presents variations of $Y'$ in the domain due to diffusion, driven by the initial spatial gradient of the stratified n-dodecane. Here, we assume that the impact of reaction on the diffusion process is negligible. Diffusion of the stratified n-dodecane can be formulated in 1d via diffusion equation under constant density ($\rho$) and constant mass diffusivity ($D$) assumptions and negligible convection,

$$\frac{\partial Y'}{\partial t} = D \frac{\partial^2 Y'}{\partial x^2}. \qquad (7)$$

Under unity Schmidt number assumption, $\nu = D$, where $\nu$ is the kinematic viscosity. Therefore, we replace $D$ by constant $\nu$ hereafter. We consider this assumption for simplification of the theoretical analysis as for the considered mixture and initial conditions in Table 1, Schmidt number is in the order of unity. Here, the initial conditions are $Y'(x, 0) = Y'(0) \sin(kx)$, and the periodic boundary conditions give $Y'(x = 0, t) = Y'(x = \lambda, t)$. Therefore, the solution of Eq. (7) reads

$$Y'(x, t) = Y'(0) \sin(kx) \exp(-k^2 \nu t), \qquad (8)$$

from which

$$\frac{\partial Y'}{\partial x} = Y'(0) k \cos(kx) \exp(-k^2 \nu t). \qquad (9)$$

The maximum of Eq. (9) is

$$\left[\frac{\partial Y'}{\partial x}\right]_{max} \simeq Y'(0) k \exp(-k^2 \nu t). \qquad (10)$$

In the present model, the maximum value above is used to calculate $S_{ign}$. We note that in this model problem, essential physics/chemistry occur at time-scales which are in the order of IDT. Substituting Eqs. (6) and (10) into Eq. (5) and setting $t = \tau$ gives

$$S_{ign}(t = \tau) \simeq \frac{1}{\alpha Y'(0) k} \exp(k^2 \nu \tau). \qquad (11)$$

Consequently, substituting $S_{ign}$ from Eq. (11) into Eq. (3) gives

$$\beta = C_\beta S_l \alpha Y'(0) k \exp(-k^2 \nu \tau). \qquad (12)$$

Considering $k = 2\pi/\lambda$, and by rearranging Eq. (12), the analytical relationship between $\lambda$ and $Y'(0)$ becomes

$$Y'(0) = \frac{\beta \lambda \exp(4\pi^2 \nu \tau/\lambda^2)}{C_\beta S_l \alpha 2\pi}. \qquad (13)$$

It should be mentioned that in the theoretical analysis above, effects of compression due to the first initial ignition is ignored. This effect might lead to slight convection in the domain which disturbs the sinusoidal curve. As this effect takes place after the first ignition, it is ignored in the theoretical analysis to allow for an analytical solution. However, this effect has been taken into account in the following 1d numerical simulations where the validity of the assumption is confirmed.

Figure 3 demonstrates theoretical presentation of the regime diagram for case 1 for the range of $\lambda$ and $Y'(0)$ using Eq. (13) when $0.01 < \beta < 100$ and assuming $S_l = 0.34$ m/s and $C_\beta = 1$. It is noteworthy that $C_\beta$ is a fitting coefficient which is always close to unity (here, $C_\beta = 1$). In the present study, $\beta = 1$ serves for the analytical borderline between deflagrative and spontaneous propagation modes, and it is called the $\beta$-curve hereafter. The values $\beta > 1$ ($\beta < 1$) are interpreted as deflagrative (spontaneous) propagation mode where $S_{ign} < S_l$ ($S_l < S_{ign}$). Normalized $Y'(0)$ is represented by $Y'_{norm} = Y'(0)/[Y'(0)]_{max}$, while wavelength is normalized by $\lambda_{diff}$ as defined in the following. Considering the maximum value of Eq. (8), $Y'(t) = Y'(0) \exp(-k^2 \nu t)$, and assuming that $\ln(Y'(\tau)/Y'(0)) = 1$, i.e. a quick damping in the order of $1/e$, the diffusion length-scale can be defined as $\lambda_{diff} = 2\pi \sqrt{\nu \tau_1}$, ($\lambda_{diff} = 0.659$ mm in case 1). Accordingly, we define $k_{diff} = 2\pi/\lambda_{diff}$, which is used to non-dimensionalize the physical domain in remainder of the paper.

Considering $\lambda$ as the dominant length-scale in this system, another consistent viewpoint for presentation of this theoretical analysis is through defining time-scales based on the propagation speeds, $\tau_{sp} = \frac{\lambda/2}{S_{ign}}$, and $\tau_{fp} = \frac{\lambda/2}{S_l}$. When $\tau_{fp} \ll \tau_{sp}$ ($\tau_{sp} \ll \tau_{fp}$), flame (spontaneous) propagation is dominant, and $\tau_{fp} \simeq \tau_{sp}$ prescribes the borderline between the two modes. When convection due to turbulence is considered, a third time-scale, convection time-scale, comes into play which is further discussed in the following section.

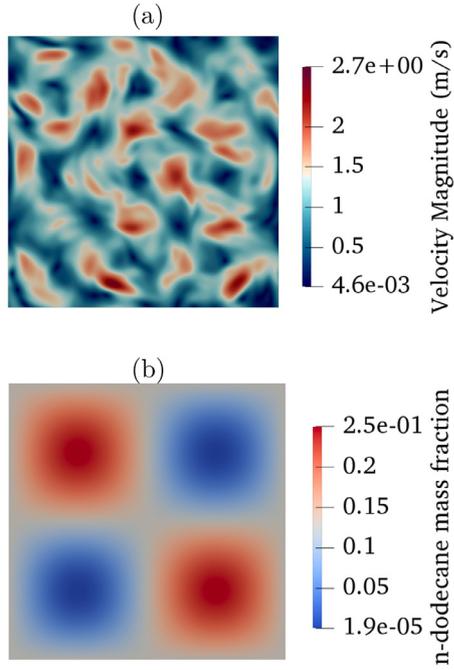

**Fig. 4.** An example of the initialization fields for: (a) velocity magnitude, and (b) n-dodecane mass fraction, in 2d simulations. The domain size is $\lambda \times \lambda$, the initial vortex size is $0.1\lambda$, while the stratification length-scale is $\lambda/2$.

**Table 3**
Convection time-scales ($\tau_c = l_{turb}/U_{turb}$) for the three different levels of turbulence. Turbulence length-scale is constant, $l_{turb} = 0.1\lambda$.

|  | Low turbulence | Medium turbulence | High turbulence |
| --- | --- | --- | --- |
| $\tau_c$ (ms) | 0.5 | 0.1 | 0.01 |
| $\tau_c/\tau_1$ | 0.45 | 0.09 | 0.009 |

### 3.5. Simulation setup in 2d

Convection due to turbulence expedites the mixing process. In order to qualitatively assess turbulence effects on modes of combustion, 2d simulations are performed for selected points in the deflagrative zone of the regime diagram. Although we are aware of the non-realistic features of 2d turbulence, such a model problem serves for the intent of this study to account for convection.

To initialize 2d simulations, turbulence is generated following the procedure described by Vuorinen and Keskinen [58] using Taylor–Green–Vortex (TGV) structure, wherein the vortex input arguments $x$ and $y$ are non-linearly perturbed in order to facilitate a fast transition to a chaotic state. Figure 4(a) depicts an example of the initial field for velocity magnitude for one of the 2d cases considered in this study. First, we let the flow field evolve from such a non-linear initial condition into a state described by a characteristic vortex size, $l_{turb} = 0.1\lambda$, and turbulent velocity, $U_{turb}$. In this study, $U_{turb}$ is chosen so that the convection time-scale $l_{turb}/U_{turb} \leq \tau$, c.f. Table 3. Second, we assume that methanol/air mixture is premixed ($\phi_{methanol} = 0.8$) while n-dodecane stratification is initialized with sinusoidal profiles in 2d,

$$Y_{ndd}(x, y) = Y_{ndd}^{ave} + Y'(0) \sin(kx) \sin(ky). \quad (14)$$

Figure 4 (b) shows an example of the initial n-dodecane mass fraction for one of the 2d cases considered in this study.

Consistent with the 1d simulations, in 2d simulations, the stratification length-scale $\lambda$ defines the domain length. In order to emulate three different turbulence levels, the simulations are set up so that $l_{turb} \approx 0.1\lambda$ while $U_{turb}$ is a free parameter controlling the convection time-scales ($\tau_c = l_{turb}/U_{turb}$), which are summarized in Table 3. In the following 2d simulations, the value of $\tau_c$ is selected so as to account for fast and slow convective effects, see Table 3. Here, the key parameter is the ratio $\tau_c/\tau$, denoting the relative dominance between mixing and chemistry.

In the 2d simulations, the $\beta$-curve is assessed only for case 1. Again, uniform initial temperature and pressure are considered, i.e. T=770 K and p = 50 bar. For each selected point on the regime diagram, the domain size in x- and y-direction is equal to the sinusoidal function wavelength ($\lambda$). Periodic boundary conditions are considered in both directions which represent the constant-volume condition. The number of gridpoints should satisfy both flow field threshold – i.e. in the order of the smallest turbulence structure – and chemical kinetics threshold – i.e. at least 20 gridpoints along the size of the laminar flame thickness. Here, the strictest spatial requirement is set by the flame thickness. The mesh size used herein is $\approx 1$ $\mu$m in all presented 2d case studies.

We note that the first ignition kernel always forms at the point with the highest n-dodecane concentration regardless of the domain size as consistent with results in Fig. 1(b). Moreover, the periodic boundary conditions in OpenFoam have been tested and compared with a pseudo-spectral Navier–Stokes code and the results are demonstrated to be consistent [58].

## 4. Numerical results and discussions

### 4.1. Phase diagrams using 1d numerical simulations

Figure 5 demonstrates the regime diagram generated for case 1 based on 150 1d simulations. A similar regime diagram is obtained for case 2, but it is not shown here for brevity. In this figure, the grayscale plot is output of 1d numerical simulations, while the curves demonstrate the theoretical $\beta$-curve borderlines provided earlier, as further explained in the following.

For each 1d numerical simulation, the criterion to distinguish between deflagrative and spontaneous ignition front propagation modes is based on the observation of temperature difference in the domain at each time instant and recording the highest temperature difference among all time instants in a simulation, denoted as $\Delta T_{max}$ hereafter. The grayscale shown in Fig. 5 is based on how close $\Delta T_{max}$ is to a threshold temperature difference, denoted as $\Delta T_{th}$. The white region corresponds to a situation with hot flame front and cool unburned gas during a simulation. In contrast, the black region corresponds to a situation with similar ignition throughout the entire domain. Here, $\Delta T_{th}$ is defined based on the reactants and products temperature difference in the steady-state solution of 1d laminar premixed flame. According to Table 2, for methanol/air laminar premixed flame at conditions of case 1 with unburned temperature at $T_u \approx 800$ K, we found $T_b \approx 1800$ K. In this respect, $\Delta T_{th} = 900$ K has been considered for case 1. Accordingly, for each point on the regime diagram, $\Delta T_{max} \geq \Delta T_{th}$ is interpreted as the deflagrative mode, and $\Delta T_{max} \ll \Delta T_{th}$ is interpreted as the spontaneous mode of combustion.

It is noteworthy that the precise choice of $\Delta T_{th}$ does not significantly influence the results shown on the regime diagram. In the present study, values in the range 800–1100 K were all providing qualitatively similar results. The idea behind using $\Delta T_{th}$ as the criterion in the regime diagram is that in the steady-state solution of planar 1d laminar premixed flame, $\Delta T = T_b - T_u$ is a unique characteristic of the flame under those conditions. In the transient problem of our interest in this study, if at any time instant a similar characteristic is observed, we interpret it as flame formation. Otherwise, in a homogeneous spontaneous ignition, $\Delta T$ is much smaller than $\Delta T_{th}$ at all time instants.

Following the theoretical analysis provided before, Eq. (13) has been utilized to mark the theoretical borderline between the two combustion modes ($\beta = 1$) in Fig. 5. As this is a transient prob-lem, the flame speed varies dynamically and does not pose an exact value. Therefore, three different values of $S_l$ have been used to plot the borderline; $0.35 \pm 0.2$ m/s for case 1 according to Table 2. Here, unity $C_\beta$ is assumed. The regime diagram presented in Fig. 5 validates the theoretical analysis and the constructed theoretical regime diagram (Fig. 3) presented before. Next, further analysis of the 1d model is provided.

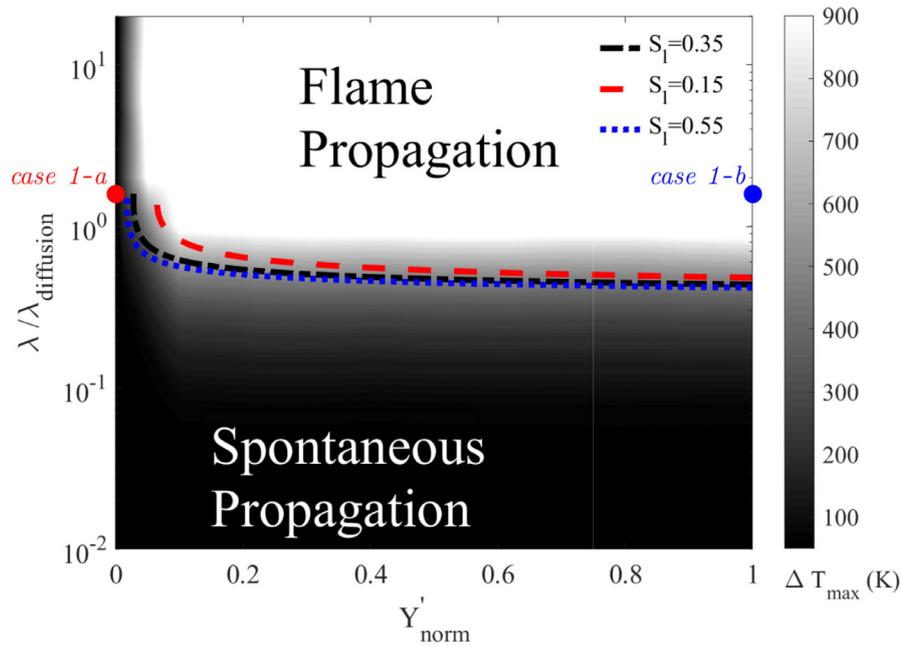

**Fig. 5.** Regime diagram based on the 1d numerical simulations for case 1 ($P = 50$ bar, $T = 770$ K). Black and white regions represent spontaneous and deflagrative zones, respectively. Dotted lines display the $\beta$-curve, Eq. (13), with $S_l = 0.35 \pm 0.2$ m/s as input ($C_\beta = 1, \beta = 1$). Here, case 1-a ($\lambda/\lambda_{diff}$=1.5, $Y'_{norm}$=0) and case 1-b ($\lambda/\lambda_{diff}$=1.5, $Y'_{norm}$=1) are selected as examples of spontaneous and deflagrative propagation, respectively.

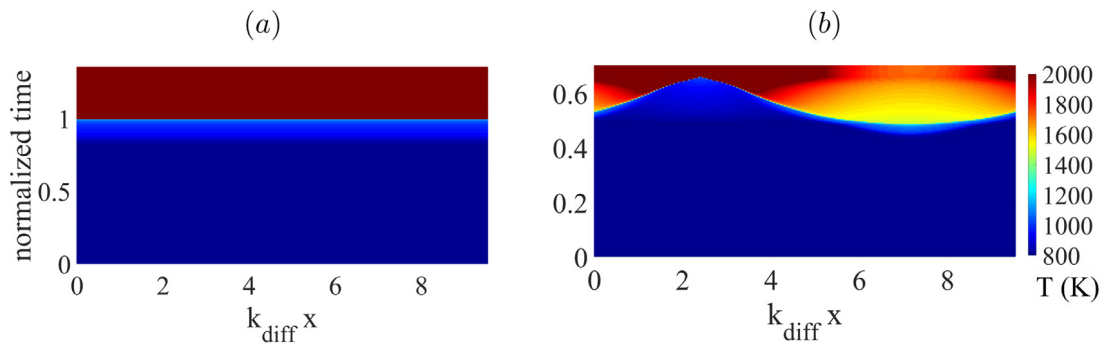

**Fig. 6.** Temperature in normalized (*x*, time) space for a point on: (a) the spontaneous zone of the regime diagram (case 1-a), and (b) the deflagrative zone of the regime diagram (case 1-b).

### 4.2. Exploration of the deflagrative mode in 1d

#### 4.2.1. Temporal evolution of ignition and flame fronts

To improve understanding on the combustion modes, two simulated points in different zones of the regime diagram are selected: one point in the spontaneous propagation region (case 1-a) and another point with deflagrative propagation dominance (case 1-b). These two points are marked in Fig. 5, and they have identical stratification wavelengths, $\lambda/\lambda_{diff}$=1.5 (1 mm), but different amplitudes; $Y'_{norm} = 0$ in case 1-a and $Y'_{norm} = 1$ in case 1-b.

Figure 6 demonstrates temperature in (*x*, time) space for case 1-a and case 1-b. Time has been normalized with IDT of the homogeneous reactor combustion for case 1, $\tau_1 = 1.1$ ms, as reported in Table 2. Moreover, domain length (*x*) is normalized by $k_{diff}$ introduced earlier in Section 3.4. In the presented cases $k_{diff} \gg 1$, indicating that the stratification length-scale is larger than the diffusion length-scale. For case 1-a, Fig. 6(a), initial ignition occurs at $t \approx \tau_1$ which is consistent with the results from Cantera for the same mixture and initial conditions. In this case, right after the IDT, the whole mixture ignites almost simultaneously, which implies spontaneous combustion. For case 1-b, Fig. 6(b), however, the first ignition kernel forms at the location of the highest n-dodecane concentration, at $k_{diff}x \approx 8$ and time $\approx 0.49\tau_1$ as expected for the richer side of the mixture according to Fig. 1(b). In contrast to case 1-a, other points of the domain ignite sequentially. In particular, as time marches after the first ignition in Fig. 1(b), the kernel grows and the temperature difference in the domain increases gradually until all reactants in the domain ignite. In the following, the structure of the reaction front in case 1-b is further analyzed.

Figure 7 depicts spatial evolution of the reaction fronts in time in case 1-b (deflagrative), where profiles of temperature (a)

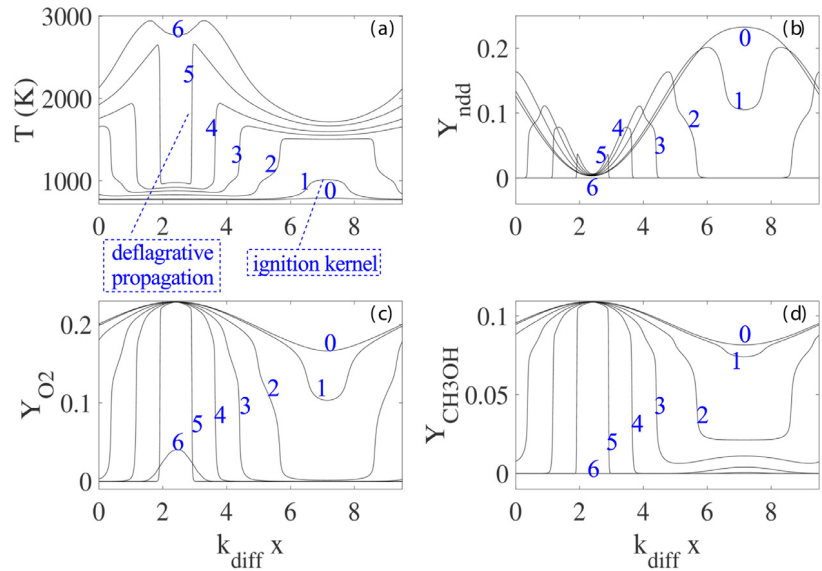

**Fig. 7.** Progress of deflagrative waves with time for case 1-b on the deflagrative zone of the regime diagram. Numbers represent temporal evolution of the isolines, starting from $0.4136\tau_1$ (no. 0) up to $0.6864\tau_1$ (no. 6) with interval $0.0455\tau_1$.

and mass fractions of n-dodecane (b), $O_2$ (c) and $CH_3OH$ (d) are demonstrated. According to the temperature profile, the early ignition kernel formed in case 1-b acts as a spark, and develops a propagating reaction front. The front propagates towards the left and right boundaries, similar to a typical premixed flame. Finally, towards the end of simulation (timeframe 6 in the figure), methanol and n-dodecane are completely burned and high-temperature products are left within the domain (at $\approx 0.69\tau_1$). It should be noted that as periodic boundaries have been used, effect of heating compression after formation of the first ignition kernel accelerates the ignition process elsewhere in the domain in the parts with lower n-dodecane concentrations. For instance, when the initial ignition kernel forms (timeframe 1 in Fig. 7), average pressure rise is $\approx 6\%$. However, the average pressure rise is $\approx 80\%$ at the time of deflagrative front formation (timeframe 5 in Fig. 7).

As noted before by Chen et al. [27], the presence of a propagating front does not necessarily imply that a normal deflagration exists. This aspect is examined by both transport budget analysis and order-of-magnitude analysis of heat release rates in what follows.

#### 4.2.2. Transport budget analysis

In the energy transport equation under laminar conditions, convection term is negligible and the balance of reaction and diffusion terms in the reaction zone indicates deflagrative propagation [59,60]. On the other hand, in spontaneous ignition, reaction dominates diffusion [61]. Energy equation budget analysis is utilized in the following to examine modes of combustion in case 1-b (c.f. Fig. 5).

Figure 8 compares diffusion and reaction terms in case 1-b at two different time instants: first, at $t = 0.5\tau_1$ when the first ignition kernel forms (a) and second, at $t = 0.6\tau_1$ when a deflagrative-like wave is present (b). The diffusion and reaction terms in each plot are normalized with respect to the corresponding maximum value of the reaction term. We note that in Fig. 8(a) and (b), the field of view is focused to the region of interest. In Fig. 8(a), it is observed that the reaction term is considerably stronger than the diffusion term, which demonstrates that subsonic spontaneous propagation is dominant. On the other hand, Fig. 8(b) demonstrates that the diffusion and reaction terms are almost in balance, which implies that a laminar premixed flame is established even in such a highly stratified and transient situation with global heat release and temperature/pressure rise.

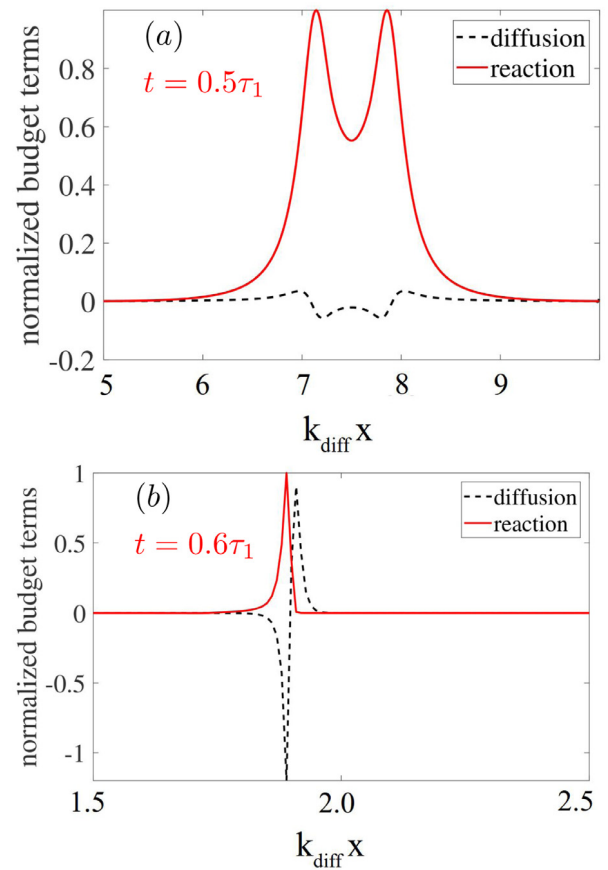

**Fig. 8.** Transport budget terms for case 1-b on the deflagrative zone of the regime diagram at the time of (a) the first ignition kernel formation, and (b) flame formation.

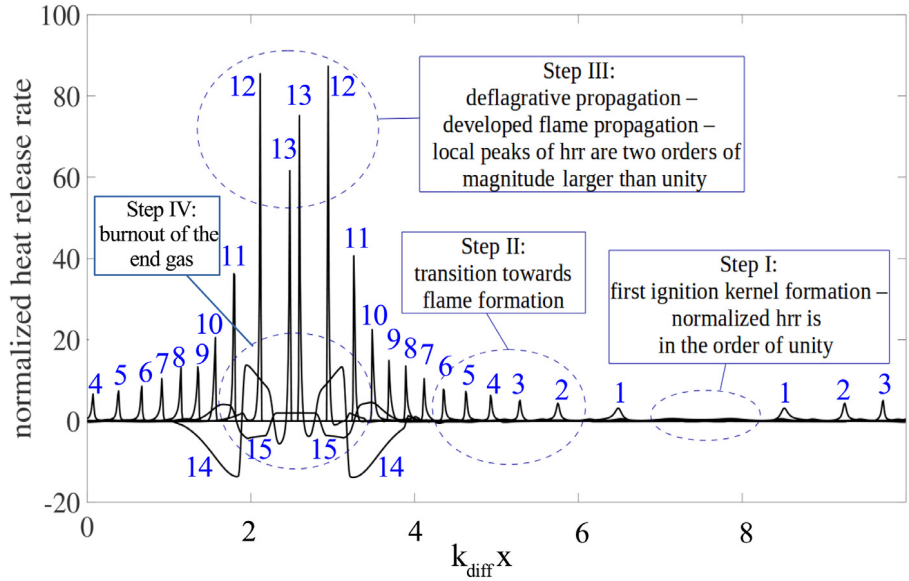

**Fig. 9.** Progress of normalized HRR with time (step I to step IV) for case 1-b on the deflagrative zone of the regime diagram.

*4.2.3. Temporal evolution of heat release rate*

Evaluation of heat release rate (HRR) magnitude and its distribution is another way to distinguish between spontaneous and deflagrative waves. In spontaneous combustion, it is expected that HRR magnitude is in the order of that in homogeneous reactor combustion and HRR should be distributed smoothly over the entire domain. In deflagrative mode, however, peaks of HRR along the flame thickness (local HRR) with comparatively higher magnitudes than that in spontaneous combustion are expected [22].

Figure 9 depicts the progress of HRR in case 1-b (c.f. Fig. 5), where HRR is normalized with maximum HRR in the 0d homogeneous reactor model combustion of the same mixture composition in case 1-b. As observed in this figure, initial HRRs (step I) are in the same order of magnitude with HRR in the homogeneous reactor case as expected during ignition. However, as time evolves, local HRR increases gradually (step II). With further time progress, developed deflagrative waves are observed (step III), similar to the flame fronts formed in Fig. 7(a). In step III, the maximum value of the locally distributed HRR is about two orders of magnitude larger than its initial value in step I, which supports the existence of deflagrative propagation in case 1-b. Finally, with burnout of the end gas, HRR decreases rapidly (step IV), completing the combustion process.

### 4.3. Convection effects and 2d simulations

In the presented theoretical regime diagram (c.f. Fig. 3), effect of convection was ignored. However, as convection expedites the mixing process, the $\beta$-curve defined earlier in Eq. (13) may not be directly applicable. In order to delineate effects of convection on the specified combustion modes, first, convection effects are incorporated to the theoretical $\beta$-expression presented earlier. Then, to qualitatively examine the validity of the theory, 2d convective flows are numerically simulated for four points in the deflagrative zone of the regime diagram. Characteristics of these points, denoted as A-D, are summarized in Table 4 and marked on the schematic regime diagram in Fig. 10. Three different levels of convection are considered at each point as earlier introduced in Table 3. The overall motivation of the chosen 2d simulation points is to demonstrate combustion mode switching from deflagrative to spontaneous due to convection.

**Table 4**
Locations of the four points selected on the deflagrative zone of the regime diagram (c.f. Fig. 10) to initiate 2d cases.

|  | Point A | Point B | Point C | Point D |
|---|---|---|---|---|
| $\lambda/\lambda_{diff}$ | 1.5 | 1.5 | 6 | 6 |
| $Y'_{norm}$ | 0.2 | 1.0 | 0.2 | 1.0 |

Effect of convection on enhanced mixing can be incorporated to Eq. (13) by assuming an effective mass diffusivity, $\nu_{eff} = \nu + \nu_{turb}$, where $\nu_{turb}$ is an estimate for increased mixing due to convection, which is dimensionally correlated with $l_{turb}U_{turb}$. We note that the unity Schmidt number assumption is used here. Consequently, Eq. (13) is modified as

$$Y'(0) = \frac{\beta\lambda \exp(4\pi^2(\nu+\nu_{turb})\tau/\lambda^2)}{C_\beta S_l \alpha 2\pi}. \quad (15)$$

In case 1, since $l_{turb} \approx 0.1\lambda$ and $l_{turb}/U_{turb} = \tau_c$, we can characterize $\nu_{turb} = l_{turb}U_{turb}$ based on $\lambda$ and $\tau_c$, which leads to $\nu_{turb} = 0.01\lambda^2/\tau_c$. Accordingly, Eq. (15) gives

$$Y'(0) = \frac{\beta\lambda \exp(4\pi^2(\nu+0.01\lambda^2/\tau_c)\tau/\lambda^2)}{C_\beta S_l \alpha 2\pi}. \quad (16)$$

Using Eq. (16), the modified $\beta$-curve at 50 bar (case 1) with convection for the three turbulence levels defined in Table 3 is presented in Fig. 10. Additionally, the $\beta$-curve at 50 bar without effects of convection is replotted using Eq. (13). It is observed that increased convection shifts the $\beta$-curve towards the deflagrative zone. This shift expands the spontaneous zone to span a broader region of the parameter space. For the lowest turbulence level ($\tau_c = 0.5$ ms), convection effects are negligible, while for the highest turbulence level ($\tau_c = 0.01$ ms), the upward shift of the $\beta$-curve is significant.

As seen in Fig. 10, according to the modified theoretical borderline in Eq. (16) with $\tau_c = 0.01$ ms (blue dotted line), the high turbulence level would switch the combustion mode for point A from deflagrative to spontaneous. This is consistent with observations of 2d numerical simulations in Fig. 10(a) and (b). The panels show time evolution of temperature fields under the high turbulence level for points A and B. In the left columns of Fig. 10(a) and (b), the first ignition kernels are established. The considered time window is 5 times longer for point B than point A. As demonstrated in Fig. 10(b), at point B, several deflagrative-like fronts are formed at different time instants and they develop at a moderate rate to eventually merge. At point A, however, the ignition kernel formed at $t \approx \tau_1$ develops and spreads throughout the domain extremely fast. It is worth noting that the initial ignition time instants at points A and B are very similar to ignition time instants of 1d cases 1-a (spontaneous) and 1-b (deflagrative), respectively.

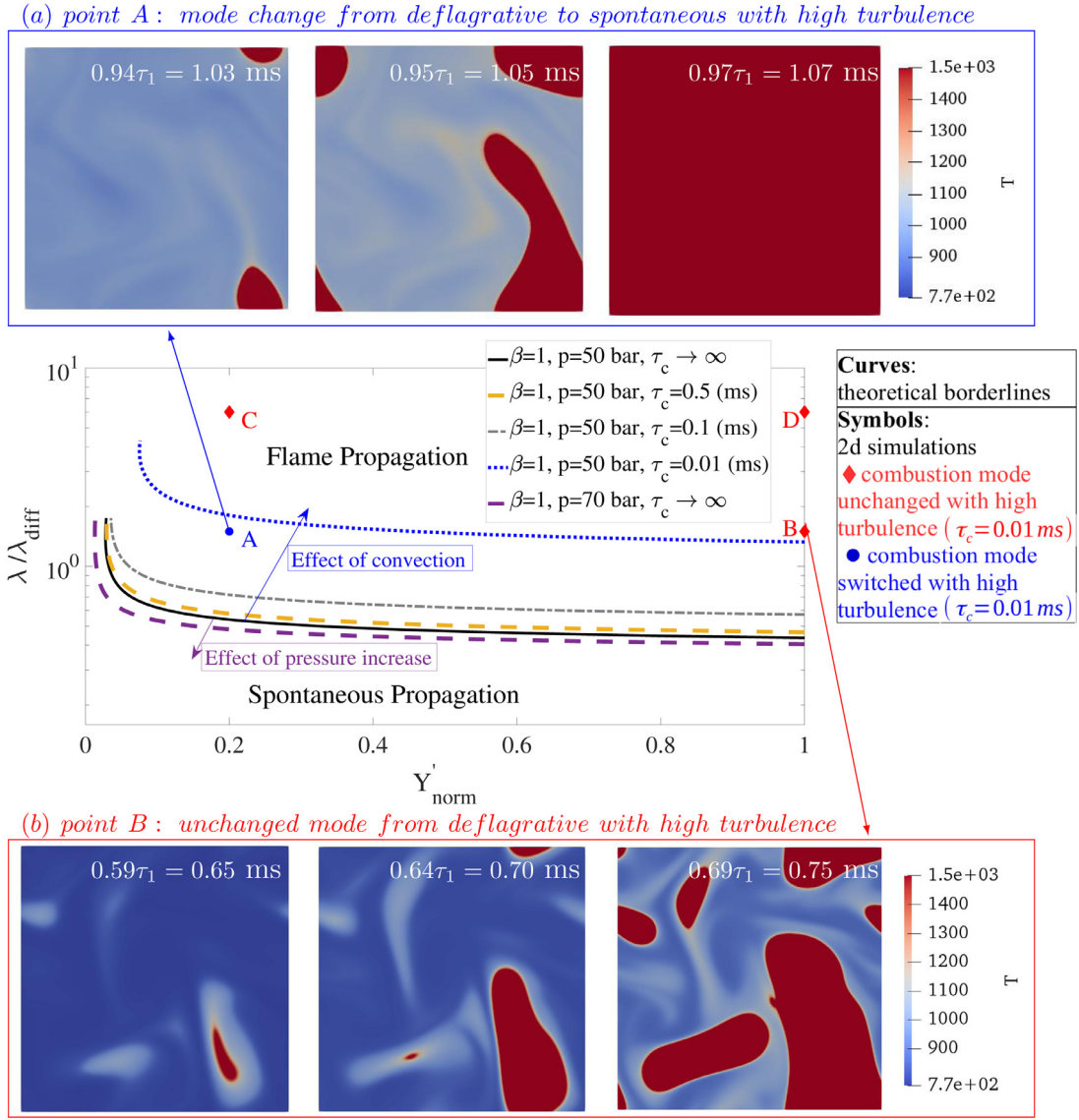

**Fig. 10.** The regime diagram (middle) summarizes effects of pressure and convection on the $\beta$-curve. Curves present theoretical borderlines. Convection with $\tau_c = 0.01$ ms changes mode of combustion only for point A: (a) time evolution for point A, and (b) time evolution for point B under high turbulence. All 2d cases are initialized with homogeneous temperature at 770 K at 50 bar. (For interpretation of the references to color in this figure, the reader is referred to the web version of this article.)

As already noted, spontaneous ignition leads to a rapid HRR, i.e. short combustion duration. In the present study, we define combustion duration as the time interval between the first ignition and the end-gas combustion time ($\tau_{end}$), denoted by $\Delta \tau$ hereafter. Here, $\tau_{end}$ is defined as the first time instant when $T_{min} > 1200$ K, $T_{min}$ being the minimum temperature within the entire domain at each time instant. Combustion duration in turbulent cases are compared with their corresponding 1d laminar cases. If $\Delta \tau$ at each point is considerably decreased compared to its corresponding 1d case, we interpret the combustion mode to switch from deflagrative to spontaneous.

Table 5 compares $\Delta \tau$ at each point for 1d numerical simulations as compared to the 2d simulations under the highest turbulence level ($\tau_c$=0.01 ms). For point A, $\Delta \tau$ has considerably ($\approx$ 70%) decreased with convection; i.e. combustion mode has changed to spontaneous due to much smaller convection time ($\tau_c$) as compared to IDT ($\tau_1$). For points B and C this difference is smaller ($\approx$ 10–20%), and for point D the difference is not distinguishable. The small discrepancies between 1d and 2d results for points B-D are mainly due to the large stratification levels (amplitude and/or wavelength) at these points. It is noteworthy that results of other 2d numerical simulations with lower turbulence levels are not shown here for brevity, but they show consistent results with the theoretical $\beta$-expression developed for different turbulence levels as presented in Fig. 10.

Finally, effect of pressure on the $\beta$-curve is elaborated in the schematic regime diagram in Fig. 10. In Fig. 10, the $\beta$-curve is plotted for case 2 ($p = 70$ bar, c.f. Table 2) according to Eq. (13). It is observed that while the effect of pressure on the $\beta$-curve is not significant, higher pressure levels broaden the deflagrative zone of the regime diagram, mainly due to the shorter IDT at the higher pressure ($\tau_2 < \tau_1$). One-dimensional numerical simulations performed for case 2 confirm this minor effect of pressure, but those results are not presented here for brevity.

Table 5
Combustion duration ($\Delta\tau$) at the four points selected on the regime diagram for the 1d simulations as compared to the 2d simulations under the high turbulence level ($\tau_c = 0.01$ ms).

| Case | Point A | | Point B | | Point C | | Point D | |
|---|---|---|---|---|---|---|---|---|
| | 1d | 2d | 1d | 2d | 1d | 2d | 1d | 2d |
| $\Delta\tau$ | $0.1\tau_1$ | $0.03\tau_1$ | $0.17\tau_1$ | $0.15\tau_1$ | $0.11\tau_1$ | $0.09\tau_1$ | $0.20\tau_1$ | $0.20\tau_1$ |

It is noteworthy that computational limitations hinder generation of modified regime diagrams based on 2d (and ideally, 3d) turbulent simulations. However, based on the analysis provided in this section, effect of convection on the combustion mode at points far from the 1d-based $\beta$-curve was insignificant. Based on the presented numerical evidence, the modified $\beta$-curve is able to predict the mode of combustion for 2d cases with non-linear mixing effects.

## 5. $\beta$-curve as a diagnostics tool

The analytical $\beta$-curve presented in this paper is proposed as a diagnostics tool to estimate modes of combustion for different fuels or fuel combinations in stratified mixtures in CI engine context. We suggest the following steps to be taken in order to have a priori understanding of the combustion mode in systems with reactivity stratification.

1. Carry out 0d homogeneous reactor simulations of combustion to estimate IDT for different mixture fractions. Consequently, estimate IDT gradient ($\alpha$) based on Eq. (6).
2. Estimate 1d laminar flame speed based on e.g., 1d numerical simulations.
3. Estimate the range of initial stratification wavelengths ($\lambda$) and amplitudes ($Y'(0)$). Then, plot $\beta$-curve based on Eq. (13) for the borderline $\beta=1$.
4. To include effect of turbulence, estimate the ranges of turbulent flow velocities and length-scales to obtain $\nu_{eff}$ based on $\nu_{eff} \propto l_{turb} U_{turb}$.
5. Plot the modified $\beta$-curve to account for convective mixing effect based on Eq. (15) for different values of $\nu_{eff}$.

We note that in an engine combustion system both modes of combustion may co-exist and the $\beta$-expression only considers certain local aspects of combustion in stratified mixtures. The focus of the present study is on understanding local combustion phenomena at relatively small length-scales. In real engines, certain relevant details such as 3d geometry and fuel injection may introduce new aspects, which are application-dependent and cannot be considered in 1d or 2d simulations. However, detailed investigation of these aspects are beyond the scope of the present study.

## 6. Conclusions

The main focus of this study was to specify whether flame fronts form and sustain in dual-reactivity mixtures - including a high reactivity fuel (HRF) and a low reactivity fuel (LRF) - under various HRF stratification levels at engine-relevant conditions. First, using an initially sinusoidal mixing problem with various wavelengths ($\lambda$) and amplitudes ($Y'$), hundreds of one-dimensional numerical simulations were carried out to distinguish between spontaneous and deflagrative modes of combustion. Second, for a one-dimensional time-dependent diffusion–reaction problem, a theoretical analysis was carried out. The analysis utilized the ideas of Zeldovich [22] and Sankaran [23] on the relative importance of ignition front propagation speed and laminar flame speed to prescribe the two modes of combustion. Third, effect of convection on the modes of combustion was examined in two-dimensional turbulent studies and finally, convective mixing effect was incorporated to the theoretical analysis. The following conclusions were made:

1. The one-dimensional theoretical analysis prescribed an analytical expression, called $\beta$-curve, for the borderline between spontaneous and deflagrative modes of combustion in ($Y'$, $\lambda$) space.
2. Using one-dimensional chemical kinetics simulations in ($Y'$, $\lambda$) space, combustion phase regime diagram was generated under engine-relevant thermodynamic conditions with almost negligible convection. It was observed that the analytical $\beta$-curve is consistent with the position of the phase border observed using chemical kinetics simulations.
3. With turbulence, convection time-scale ($\tau_c$) plays a role as compared to ignition delay time (IDT). When $\tau_c \ll$ IDT, convection expedites mixing and influences the borderline on the combustion phase regime diagram. Effect of convection was incorporated to the theoretical analysis through defining effective kinematic viscosity/mass diffusivity ($\nu_{eff} = \nu + \nu_{turb}$) under unity Schmidt number assumption, where $\nu_{turb}$ is correlated with turbulence length-scale and velocity.
4. Several two-dimensional turbulent chemical kinetics simulations were performed and it was observed that convective mixing can expand the spontaneous ignition zone by switching the combustion mode from deflagrative to spontaneous for points close to the borderline depending on their stratification levels and the turbulence intensity. Additionally, the effect of pressure on the borderline position was observed to be minor.

We note that an interesting future avenue to further develop the $\beta$-expression with convection effects is to define time-averaged effective $\lambda$, $Y'$ and IDT values over the ignition delay period, for example using the integral approach in Ref. [62]. Besides, as the presented analytical $\beta$-curve is able to provide a priori knowledge about modes of combustion, the following pathway of the research is to apply the concept to classification of different practical applications.

## Declaration of Competing Interest

The authors declare that they have no known competing financial interests or personal relationships that could have appeared to influence the work reported in this paper.

## Acknowledgments

The present study has been financially supported by the Academy of Finland (grant numbers 318024 and 297248). We would like to acknowledge Aalto University and CSC (Finnish IT Center for Science) for providing the computational resources.

## Appendix A. Validation of chemical kinetics mechanism

The purpose of this appendix is to present validation results for the Polimi reduced mechanism to be used in DF n-dodecane/methanol/air combustion. As mentioned in Section 2.2, since there is no experimental data for DF methanol/n-dodecane combustion in the literature, the validation is performed for SF conditions only. The Polimi reduced mechanism has been validated and used for SF n-dodecane/air under engine-relevant conditions in our previous computational studies of n-dodecane/methane DF combustion, in Refs. [48,63]. Here, it is only validated for SF methanol/air combustion. Due to the limited available experimental data on methanol combustion, especially at engine-relevant high pressure conditions, the mechanism is validated against other available chemical mechanisms developed for methanol as well. To elaborate on the effect of reduced reactions on the accuracy of the Polimi reduced mechanism, the detailed version [64] – called Polimi detailed hereafter – is also included in this validation analysis.

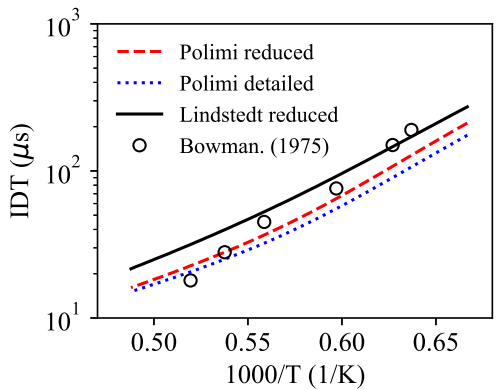

**Fig. A1.** IDT calculations under different initial temperatures for atmospheric-pressure methanol/air combustion using different chemical mechanisms as compared with the experimental data in Ref. [67].

Table A.1 lists specifications of the chemical mechanisms used for validation of the Polimi reduced mechanism. Lindstedt reduced mechanism [66] has been previously validated against the experimental data of Bowman [67] for atmospheric pressure methanol combustion in 0d homogeneous reactor and against the experimental data of Egolfopoulos et al. [68] for atmospheric pressure methanol combustion in 1d laminar premixed flames. In the current study, it is used for validations under atmospheric pressure conditions against the Polimi reduced mechanism. For validations under high pressure conditions, Klippenstein reduced mechanism, [65], is utilized along with the available experimental data for high pressure methanol/air combustion by Burke et al. [69]. We note that the Klippenstein mechanism is the improved version of the mechanism by Juan et al. [70] for high pressure conditions.

Figure A.1 presents calculations of IDT for methanol/air combustion in a 0d homogeneous reactor at atmospheric pressure with $\phi = 0.75$ using different chemical mechanisms as compared with the experimental data by Bowman et al. [67]. It is observed that the Polimi reduced reproduces a similar trend with that of the Lindstedt reduced mechanism. In addition, IDT values calculated using the Polimi reduced are consistent with the experimental data, especially under higher temperature values. For lower temperatures, although the Lindstedt mechanism results are more con-

**Table A1**
Characteristics of the utilized chemical mechanisms for validation.

| Mechanism | No. of species | Reaction steps |
|---|---|---|
| Polimi reduced [52] | 96 | 993 |
| Polimi detailed [64] | 451 | 17,848 |
| Klippenstein 2011 [65] | 21 | 93 |
| Lindstedt reduced [66] | 32 | 167 |

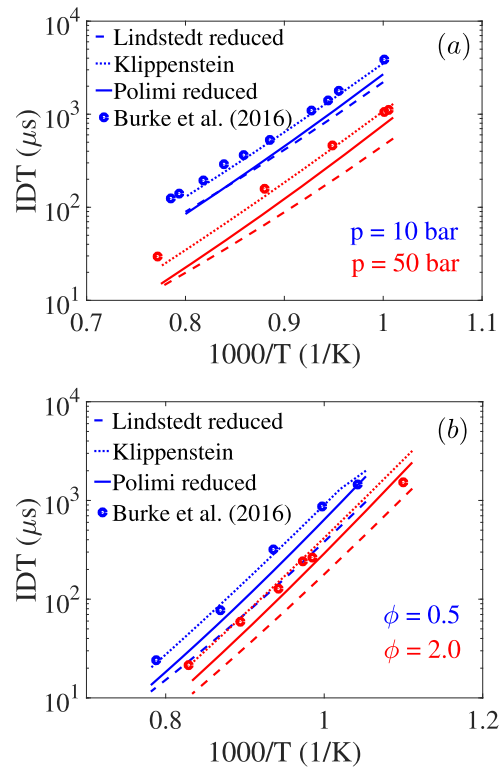

**Fig. A2.** Effects of: (a) pressure at $\phi = 1.0$, and (b) equivalence ratio at $p = 50$ bar, on IDT at different initial temperatures for methanol/air combustion using different chemical mechanisms as compared with the experiments in Ref. [69].

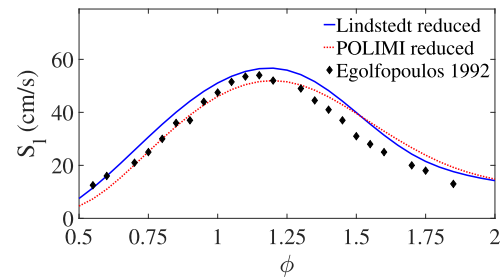

**Fig. A3.** Laminar flame speed against equivalence ratio for premixed methanol/air combustion under atmospheric conditions using different chemical mechanisms as compared with the experimental data in Ref. [68].

sistent with the experimental data, the Polimi reduced results do not deviate considerably from the experimental data.

Figure A.2 presents effects of pressure at stoichiometric conditions (a), and effects of equivalence ratio at pressures around 50 bar (b), using different chemical mechanisms as compared with the experimental data by Burke et al. [69]. While the Klippenstein mechanism, specifically developed for high-pressure methanol combustion, can reproduce the experimental data very well, the Polimi reduced provides more precise results compared with the Lindstedt reduced mechanism, and its deviations from the experimental data are small. This difference is smaller at higher pressures and lower temperatures under the considered range of pressure and equivalence ratio.

Figure A.3 demonstrates 1d premixed laminar flame speed results calculated in atmospheric pressure for stoichiometric methanol/air combustion similar to conditions considered in experiments by Egolfopoulos et al. [68] and as presented in Ref. [66]. It is observed that for $0.7 < \phi < 0.9$, the typical range of equivalence ratio in engines, the Polimi reduced is even more consistent with the experimental data compared with the Lindstedt reduced mechanism, which is specifically developed for methanol combustion.